\documentstyle[12pt,epsf]{article}
\begin{document}

%

\pagestyle{plain}

\title{Physics at the B Factories}
\author{ T.E. Browder, University of Hawaii at Manoa}
\date{}
\maketitle

\medskip
\centerline{(To appear in the Proceedings of 
From the Smallest to the Largest Distances,} 
\centerline{a conference
in honor of  Tranh Thanh Van in Moscow, Russia).}

\begin{abstract}

We review recent progress at the two $e^+ e^-$ B factories.
The first measurement of CP violation and the prospects
for measuring all the angles of the unitarity triangle are discussed. 

\end{abstract}
\bigskip

\def\Journal#1#2#3#4{{#1} {\bf #2}, #3 (#4)}

\def\NCA{\em Nuovo Cimento}
\def\NIM{\em Nucl. Instrum. Methods}
\def\NIMA{{\em Nucl. Instrum. Methods} A}
\def\NPB{{\em Nucl. Phys.} B}
\def\PLB{{\em Phys. Lett.}  B}
\def\PRL{\em Phys. Rev. Lett.}
\def\PRD{{\em Phys. Rev.} D}
\def\ZPC{{\em Z. Phys.} C}

\def\st{\scriptstyle}
\def\sst{\scriptscriptstyle}
\def\mco{\multicolumn}
\def\epp{\epsilon^{\prime}}
\def\vep{\varepsilon}
\def\ra{\rightarrow}
\def\ppg{\pi^+\pi^-\gamma}
\def\vp{{\bf p}}
\def\ko{K^0}
\def\kb{\bar{K^0}}
\def\al{\alpha}
\def\ab{\bar{\alpha}}
\def\be{\begin{equation}}
\def\ee{\end{equation}}
\def\bea{\begin{eqnarray}}
\def\eea{\end{eqnarray}}
\def\CPbar{\hbox{{\rm CP}\hskip-1.80em{/}}}

\bibliographystyle{unsrt}    

\section{High Luminosity $e^+ e^-$ B factories}
\subsection{Introduction}\label{subsec:prod}

Many important discoveries in high energy physics were 
announced in the Rencontres de Moriond series. For example,
at the 1987 Rencontres de Moriond in a presentation by H. Schroeder,
the ARGUS collaboration described the surprising discovery of large
$B_d-\bar{B_d}$ mixing\cite{ARGUS}. The implications of this observation 
were important for the experimental program on  CP violation. 
As a result, the theoretical
proposal of Bigi, Carter and Sanda\cite{carter} that there would be large
CP asymmetries in hadronic $B$ decays to CP eigenstates became
experimentally feasible. This initiated the program which
would eventually lead to the plans for the high luminosity $e^+e^-$
B factories.

Two international high luminosity $e^+ e^-$ $B$ factories 
have now been completed and are collecting data in the
United States (PEPII/BABAR) and Japan (KEKB/BELLE)
\cite{BABAR_nim},\cite{BELLE_nim}. 
The goal of these experiments
is to measure CP asymmetries in the $B$ system and test the
foundations of the Standard Model. 

The accelerators
were commissioned with remarkable speed starting in late 1998. 
The experiments starting physics data taking in 1999.
By the time of the XXXVI Rencontres
de Moriond in winter 2001, there were hints (at the 1.7$\sigma$ level)
of CP violation from both BABAR and BELLE. 
In the summer of 2001, these two experiments
announced the observation of the 
first statistically significant signals for CP violation 
outside of the kaon system.

\subsection{CP Violation Induced by Mixing}

In 1973, Kobayashi and Maskawa (KM) first proposed a model
where $CP$ violation is
incorporated as an irreducible complex phase in the
weak-interaction quark mixing matrix~\cite{KM}.
The idea, which was presented at a time when only the $u$, $d$ and
$s$ quarks were known to exist,  was remarkable because
it required the existence of six quarks.
The subsequent
discoveries of the $c$, $b$ and $t$ quarks, and the compatibility
of the model with the $CP$ violation observed in the neutral $K$
meson system led to the 
incorporation of the KM mechanism into the Standard Model,
even though it had not been conclusively tested experimentally.

The unitarity of the CKM matrix implies that the existence of three
measurable phases. In the ``Nihongo'' convention, these are denoted
\begin{equation}
\phi_1\equiv arg
\left( \begin{array}{c}
-\frac{V_{cd}V^*_{cb}}{V_{td}V^*_{tb}}
\end{array} \right) 
,~~ \phi_2\equiv arg
\left( \begin{array}{c}
-\frac{V_{ud}V^*_{ub}}{V_{td}V^*_{tb}}
\end{array} \right)
,~~ \phi_3\equiv arg
\left( \begin{array}{c}
-\frac{V_{cd}V^*_{cb}}{V_{ud}V^*_{ub}}
\end{array} \right).
\end{equation}
while at SLAC these angles are usually 
referred to as $\beta, \alpha$ and $\gamma$, respectively.

A non-zero value of $\phi_{CP}$ results in 
the time dependent asymmetry 
\begin{equation}
 A_f =\frac{R(B^0\to f_{CP} )-R(\bar{B}^0\to f_{CP})}
{R(B^0\to f_{CP} )+R(\bar{B}^0\to f_{CP} )} 
=\xi_f \sin 2\phi_{CP} \cdot \sin (\Delta m \cdot (t_2 \pm t_1) ), 
\end{equation}
where $\xi_f$ is the CP eigenvalue ($\pm 1$),
$\Delta m$ denotes the mass difference between the two $B^0$ mass
eigenstates and $t_1$ and $t_2$ are the proper time
for the tagged-$B$ and $CP$ eigenstate decays, respectively. 
The $+$ sign corresponds to the case where the $B^0$ and $\overline{B^0}$ 
are in an even L orbital angular momentum state, the $-$ sign 
obtains for odd L states.  A determination of $A_f$ thus provides
a measurement of $\sin 2\phi_{CP}.$

We note that due to the restrictions of quantum mechanics,
time integrated asymmetries at the $\Upsilon(4S)$ resonance 
(which corresponds to the $-$ sign in the above equation) are identically
zero. Therefore, one must make time dependent measurements. Since the 
pairs of B mesons are produced nearly at rest in the usual arrangement
at threshold, the $\Upsilon(4S)$ center of mass frame
must be boosted. This is accomplished by the use of beams with
asymmetric energies.
For example at KEK-B, $\beta\gamma\sim 0.43$, 
(at PEP-II, $\beta\gamma\sim 0.56$)
and the typical $B$
meson decay length is dilated from $\sim 20\mu$m
to $\sim 200 \mu$m, which is measurable with solid-state vertex detectors
close to the interaction point. 
At the symmetric CESR storage ring which uses equal energy beams, these
measurements are not possible. 

\subsection{Accelerators and Detectors}

The $e^+ e^-$ $B$ factory experiments require substantial
improvements in accelerator technology to achieve luminosities in 
excess of $3\times 10^{33}/{\rm cm}^2/{\rm sec}$ and data samples
with integrated luminosities of {\cal O}($100$ fb$^{-1}$)
as well as some improvements in detector technology. 
The properties of the B factory accelerators are summarized in Table~1.
The properties of the individual bunches are not dramatically different
from what had been achieved at conventional storage rings
such as CESR.
Instead to obtain high luminosity, 
the asymmetric B factories use double storage rings and
very large numbers of bunches. This is a domain of accelerator operation
which had not been fully explored previously 
and to avoid instabilities
involving coupling of bunches requires the operation of complex
feedback systems. 

To minimize
parastic collisions between incoming and outgoing bunches, 
KEK-B chose the crossing angle approach while SLAC
takes advantage of the energy asymmetry and uses magnetic separation
to achieve this separation. A small crossing angle of $ \pm 2.1$ mrad
had been successfully used at CESR to achieve high luminosity.
 KEK-B employs a somewhat larger $\pm 11$ mrad crossing angle. 
There is still some worry that
coupled oscillations of the longitudinal and transverse degrees of the
beam (synchro-betatron oscillations) may be excited, so KEK-B continues
R+D to develop special RF cavities to rotate the bunches so that they collide
head on at the interaction point. The magnetic separation approach gives
rise to somewhat larger experimental backgrounds and imposes tight
constraints on the engineering of the interaction region. 

The performance achieved by both accelerator complexes in a very
short time is quite remarkable. 
Both started physics running in 1999. By the end
of the summer of 2001, PEP-II had integrated about 40 fb$^{-1}$ while
KEK-B had integrated 33 fb$^{-1}$. The peak luminosity for
PEP-II was $3.9 \times 10^{33}$/cm$^2$/sec while KEK-B had achieved 
$4.5 \times 10^{33}$/cm$^2$/sec. In the case of KEK-B, the beam
currents are still far below the design values. (Also note that the
operating parameters currently used  are constantly changing
and somewhat different from those in the design (Table 1)).

The experimental measurements of CP violation
require tagging of the $B$ flavor, 
 measurement of the time dependence of the decay and high momentum
particle identification. As shown in Table~2,
in contrast to the accelerators, the detectors are quite similar with 
the exception of the 
particle identification systems\cite{BABAR_nim},\cite{BELLE_nim}.

\begin{center}
\begin{tabular}{|lc|cc|cc|} \hline
\multicolumn{6}{|c|}{ Table 1: 
Design Parameters for B-Factory Accelerators} \\ \hline
\multicolumn{2}{|c|}{ } &\multicolumn{2}{|c|}{KEK-B}&
\multicolumn{2}{|c|}{SLAC}  \\ 
\hline
  &  & LER & HER & LER & HER   \\ 
Energy& $E$(GeV) & 3.5 & 8.0 & 3.1 & 9.0   \\
Luminosity& $L\rm (cm^{-2}s^{-1})$  &\multicolumn{2}{|c|}{$1\times 10^{34}$}&
\multicolumn{2}{|c|}{$3\times 10^{33}$}  \\
Collision mode&  &\multicolumn{2}{|c|}{$\pm 11 mr$ (crab)} &
\multicolumn{2}{|c|}{Head-on}  \\ 
Circumference& $C$(m) &\multicolumn{2}{|c|}{3018} &\multicolumn{2}{|c|}{2199} 
 \\
Beta Function& $\beta^*_x/\beta^*_y$(cm)  & 100/1 & 100/1 & 37.5/1.5 & 75/3 
 \\
Tune shift& $\xi_x/\xi_y$ & \multicolumn{2}{|c|}{0.05/0.05} &
\multicolumn{2}{|c|}{0.03/0.03}   \\
Emittance & $\epsilon_x/\epsilon_y$(nm) & 19/0.19 & 19/0.19 &
64/2.6 & 48.2/1.9  \\
Energy spread & $\sigma_E/E(10^{-4}$) 
& 7.7 & 7.2 & 9.5 & 6.1   \\
Total Current & $I$(A) & 2.6 & 1.1 & 2.14 & 0.98
 \\
No. of Bunches & $N_B$ & \multicolumn{2}{|c|}{5120} &
\multicolumn{2}{|c|}{1658}  \\
Bunch spacing & $S_B$(m) & \multicolumn{2}{|c|}{0.6} &
\multicolumn{2}{|c|}{1.26}  \\
RF frequency & $f_{RF}$(MHz) &  \multicolumn{2}{|c|}{508} &
\multicolumn{2}{|c|}{476}   \\
RF voltage & $V_c$(MV) & 22 & 48 & 9.5 & 18.5   \\
Cavity type & & ARES & super &
\multicolumn{2}{|c|}{1-cell normal}  
  \\
No. of Cavities & $N_c$ & 28 & 60 & 10 & 20    \\
\hline
\end{tabular}
\end{center}

The most challenging requirement
for the detector is the detection of short lived vertices.
The experiments at both $B$ factories use double-sided silicon strips
to deal with this requirement. These have been used previously at ALEPH,
DELPHI, and at CLEO II and III. Nevertheless, the implementation is still
far from routine. 

The other difficult 
requirement for the detector is the separation
of kaons from pions. At high momentum, this is needed to distinguish
$\bar{B}^0\to \pi^+ \pi^-$ from $\bar{B}^0\to K^-\pi^+$. The former is a
spectator $b\to u$ mode which is used to determine one of the angles in the
unitarity triangle while the latter is an example of a gluonic
penguin $b\to s g^*$ decay. Similarly, the experiments must
distinguish the Cabibbo suppressed mode $B^-\to D^0 K^-$ from the 
much more abundant and kinematically
similar Cabibbo favored mode $B^-\to D^0 \pi^-$. At lower momenta,
particle identification is essential for flavor tagging.

Two approaches for high momentum particle identification have been
implemented. Both are based on the use of Cerenkov radiation.
At BELLE, aerogel Cerenkov radiators are used. Blocks of
aerogel are readout directly by fine-mesh phototubes which have
high-gain and operate comfortably in a 1.5 Tesla magnetic field. 
Since the threshold for the aerogel is around 1.5 GeV, below this
momentum K/$\pi$ separation is carried out using high precision time-of-flight
scintillators with resolution of 95 ps. The aerogel and TOF counter
system are complemented by dE/dx measurements in the central drift
chamber. The dE/dx system provides K/$\pi$ separation around 2.5 GeV
in the relativistic rise region and below 0.7 GeV.
For high momentum kaons, an efficiency of 88\% with a
misidentification probability of 9\% has been achieved.

At BABAR, Cerenkov light is produced in quartz bars and then transmitted
by internal reflection outside of the detector through a water tank
to a large array of phototubes
where the ring is imaged. The detector is referred to by the acronym
DIRC. It provides particle identification over the full momentum
range for particles that are energetic enough to reach it.
Additional particle identification is provided by dE/dx 
measurements from the
drift chamber and from the 5-layer silicon detector system.

For tracking, both B-factory experiments use conventional drift chambers
with a low-Z helium based gas to minimize multiple scattering
and SR (synchrotron radiation) backgrounds. 

To detect photons and electrons, the $B$ factory detectors use 
large arrays of CsI(Tl) 
crystals located inside the coil of the magnet. These
were used previously with great success by the CLEO II experiment.
In BABAR and BELLE, another novel
feature is a hadron calorimeter which is used for $K_L$
and muon detection and allows the use of $B^0\to \psi K_L$ in addition to 
the CP eigenstate $B\to \psi K_s$.

\begin{center}
\begin{tabular}{|l|c|c|c|} \hline
   \multicolumn{3}{|c|}{Table 2: B Factory Detectors}\\ \hline
 Name      &  BaBar         &  BELLE               \\ 
 Logo      &  Elephant      &  Jazzy $B$         \\
           &                &                              \\ 
$\theta_{crossing}$ &  $0$ mr & $\pm 11$ mr       \\
           &                &                                \\ 
Vertex     &  5 layers      & 3 layers          \\
$r$ (cm)   &$3.2-14.4$   & $3- 5.8$         \\
           &                &                              \\ 
CDC        &  40 layers     & 50 layers           \\
$r$ (cm)   & $24- 80$     & $9- 86$          \\
           &                &                              \\ 
PID        &   DIRC+dE/dx   & Aerogel/TOF+dE/dx             \\
           &                &                              \\ 
EM Cal     &  CsI(Tl)       &  CsI(Tl)                      \\
           &                &                              \\ 
Magnet     &  1.5 T         & 1.5 T                   \\
           &                &                               \\ 
$K_L/\mu$  &  RPC           & RPC               \\
           & Linseed Oil    & Glass                        \\
           &                &                              \\ 
DAQ        & Digital        & Q-to-T into            \\
           & Pipeline       & multihit TDC        \\\hline
\end{tabular}
\end{center}

\subsection{The First Measurements of CP Violation}

The measurement requires the reconstruction of $B^0\to f_{CP}$
decays, the determination of the $b$-flavor of the accompanying (tagging)
$B$ meson, the measurement of $\Delta t$,
and a fit of the expected $\Delta t$ distribution to 
the measured distribution using a likelihood method.

BELLE reconstructs
 $B^0$ decays to the following ${CP}$ eigenstates:
$J/\psi K_S$, $\psi(2S)K_S$, $\chi_{c1}K_S$, $\eta_c K_S$ for $\xi_f=-1$  and
$J/\psi K_L$ for $\xi_f=+1$. BABAR includes all these modes
except for $\eta_c K_s$. 
The two classes ($\xi_f=\pm 1$)
should have CP asymmetries which are opposite in sign.

Both experiments also use $B^0\to J/\psi K^{*0}$ decays where
$K^{*0}\to  K_S\pi^0$.  
Here the final state is a mixture of even and odd CP.
The CP content can, however, be determined from an
angular analysis of other $\psi K^*$ decays. The CP odd
fraction is found to be small (i.e. $19\pm 4$ \% ($16\pm 3.5$) 
in the BELLE (BABAR) analysis).

\begin{figure}[htb]
\centerline{\epsfysize 3.5 truein\epsfbox{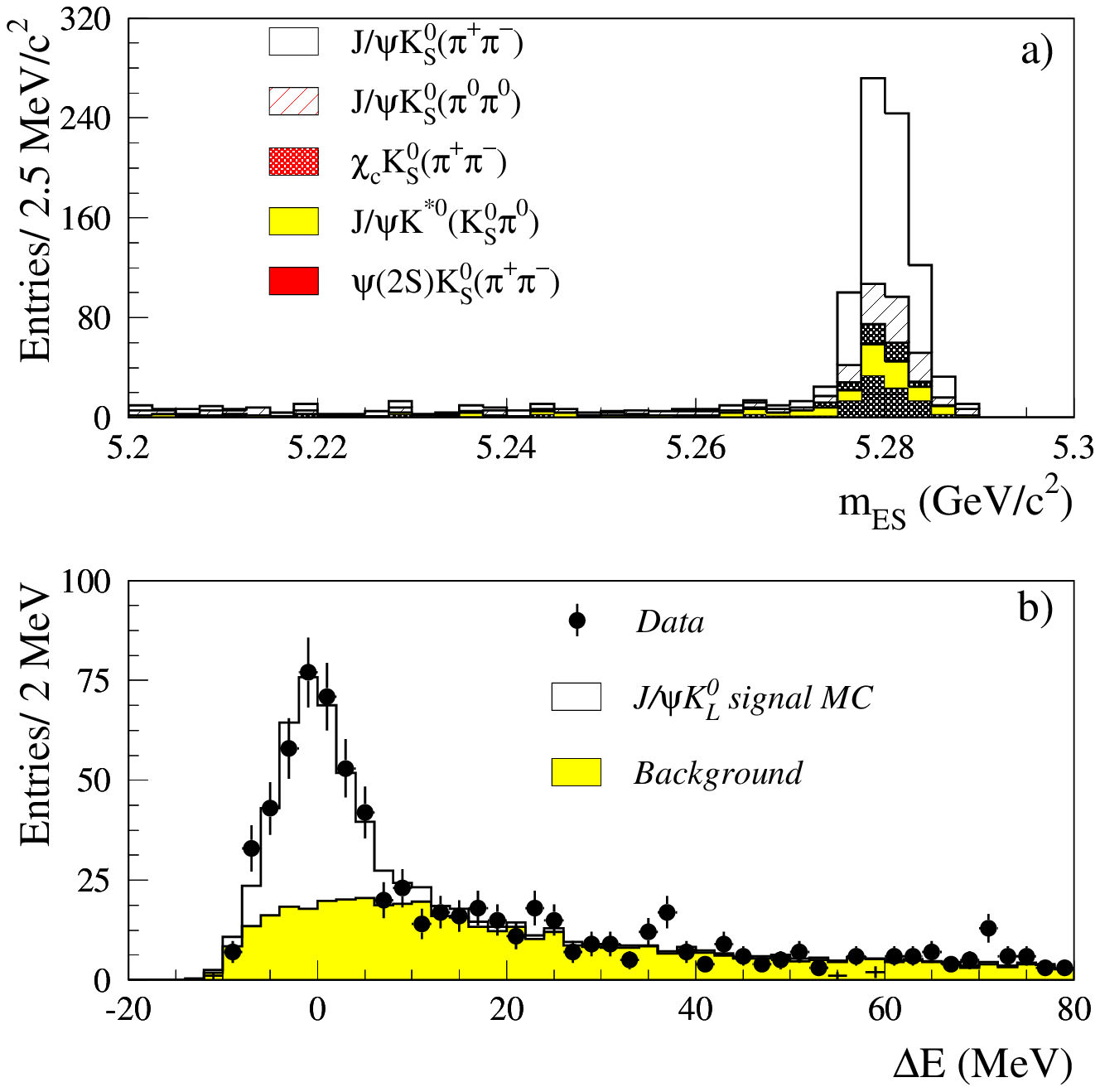}}
\caption{BABAR CP eigenstates. The upper figure shows the
beam constrained mass distribution of the fully reconstrructed
CP eigenstates. The lower figure shows the $p_B^*$ distribution
for the $B^0\to \psi K_L$ candidates.}
\label{babar_fig1}
\end{figure}

\begin{figure}[htb]
\centerline{\epsfysize 3.5 truein\epsfbox{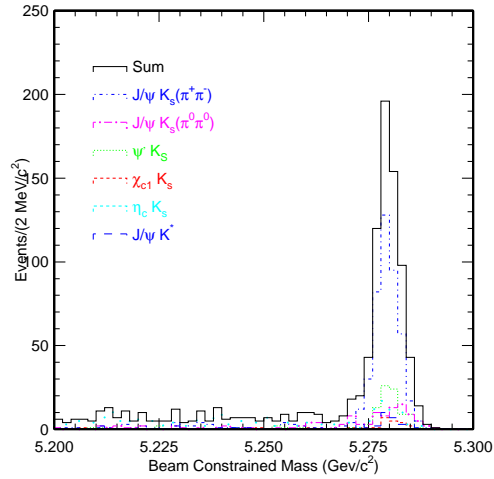}}
\centerline{\epsfysize 3.0 truein\epsfbox{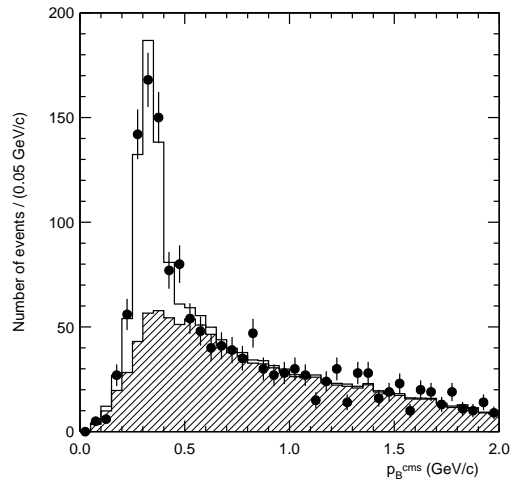}}
\caption{Belle CP eigenstates: the upper figure shows
the beam constrained mass distributions for the exclusive
modes while the lower figure shows the $p_B^*$ distribution
for the $B^0\to \psi K_L$ candidates.}
\label{belle_fig1}
\end{figure}

The CP eigenstate event samples are shown in Figs.~\ref{babar_fig1}
and \ref{belle_fig1}. The distributions of beam constrained mass
are shown for the modes other than $\psi K_L$. For the latter,
the distribution of $p_B^*$, the momentum of the B candidate in
the CM frame is shown. 
In the fully reconstructed modes,
BELLE finds 747 events over a background of 58.6.
For the $\psi K_L$ sample, there are 569
over a background of 223. The event yields from BABAR are similar.
They observe a total of 1230 events over background of 200.
It is clear that the 
CP eigenstate samples that are used for the CP violation
measurements are large and clean.

To identify
the flavor of the accompanying $B$ meson,
leptons, kaons, charged slow pions from $D^*\to D^0\pi^+$ decays, 
and energetic pions from two-body $B$ decay (e.g. $\bar{B^0}\to
D^{*+}\pi^+$) are used. 
A likelihood based method is used by BELLE. BABAR first separately
tags leptons and then kaons. If these methods fail or conflict,
two neural nets are then used
to identify the remaining tags. The figure of merit for flavor
tagging is the effective efficiency, $\epsilon_{eff}$, 
which is $\epsilon(1-2 w^2)$ summed over all tagging categories.
BELLE's method gives 
$\epsilon_{eff} = 0.270\pm 0.008^{+0.006}_{-0.009}$
while BABAR finds $\epsilon_{eff} = 0.261\pm 0.012$ for their method.

The probabilities of an
incorrect flavor assignment
are determined directly from the data 
using exclusively reconstructed, self-tagged
$B^0\to D^{*-}\ell^+\nu$, $D^{(*)-}\pi^+$,  
$D^{*-}\rho^+$  and $J/\psi K^{*0}(K^+\pi^-)$ decays.
The values of
$w_l$ are obtained from the amplitudes of the
time-dependent $B^0\overline{B}{}^0$ mixing oscillations:
$(N_{\rm OF} - N_{\rm SF})/(N_{\rm OF}+N_{\rm SF})
=(1-2w_l )\cos (\Delta m_d \Delta t)$.
Here $N_{\rm OF}$ and $N_{\rm SF}$ are the numbers of opposite and same
flavor events. These wrong-tagging rates determined from data
are then used in the CP fit for BELLE. To incorporate
this information in their analysis, BABAR performs a simultaneous
fit to the flavor tagged and CP eigenstate data samples.

The $f_{CP}$ vertex is determined using
lepton tracks  from $J/\psi$ or $\psi(2S)$ decays, 
or prompt tracks from $\eta_c$ decays.
The $f_{\rm tag}$ vertex
is determined from well reconstructed tracks not assigned to $f_{CP}$.
Tracks that form a $K_S$ are not used.
For BELLE, the typical vertex-finding efficiency and
vertex resolution (rms) for $z_{CP}$ ($z_{\rm tag}$) are
$92~(91)\%$ and $75~(140)~\mu{\rm m}$, respectively. 
Note that the resolution on the tag side includes a large 
additional contribution from charm decay.
For BABAR, the rms resolution in $\Delta z=z_{CP}-z_{tag}$
is 180$\mu$m and their vertexing efficiency is 97\%.
 Once the tag and CP eigenstate vertices are reconstructed,
the proper time is calculated from $\Delta z/\gamma \beta$.

\begin{figure}[htb]
\centerline{\epsfysize 5.5 truein\epsfbox{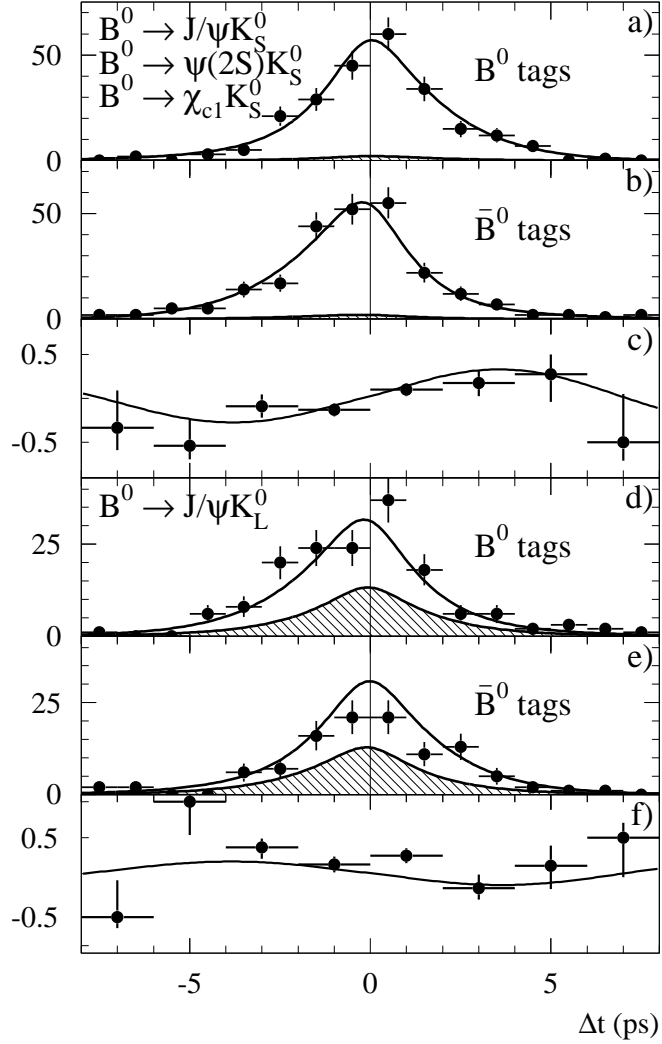}}
\caption{BABAR data on time dependent CP asymmetry.
For the fully reconstructed $CP=-1$ final states,
(a) The $\Delta t$ distributions for $B^0$ tags;
(b) The $\Delta t$ distributions for $\bar{B}^0$ tags;
the shaded area is the background, the solid curve is
the fit to the data.
(c) The time dependent asymmetry between (a) and (b);
For the $\psi K_L$ final state,
(d) The $\Delta t$ distributions for $B^0$ tags;
(e) The $\Delta t$ distributions for $B^0$ tags;
(f) The time dependent asymmetry between (d) and (e);}
\label{babar_fig2}
\end{figure}

\begin{figure}[htb]
\centerline{\epsfysize 3.5 truein\epsfbox{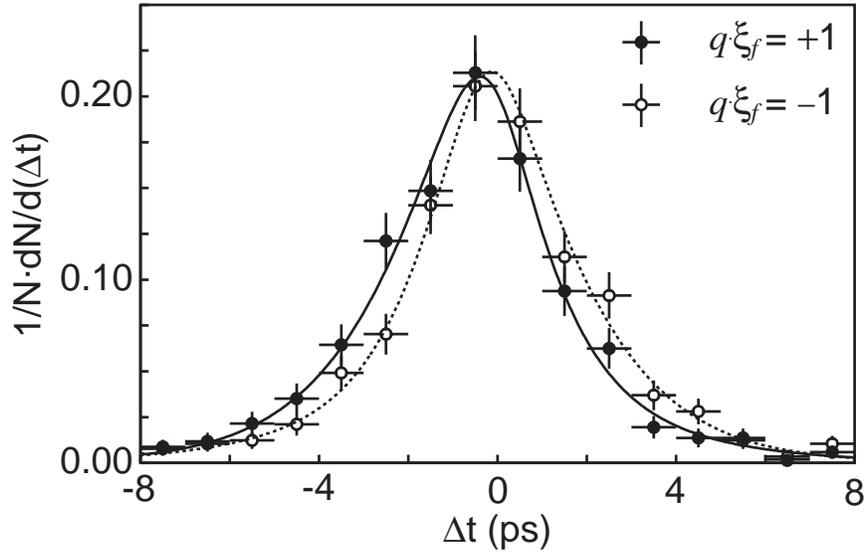}}
\caption{BELLE data on time dependent CP asymmetry.
$\Delta t$ distributions 
for the events with $q\xi_f = +1$ (solid
points) and $q\xi_f = -1$ (open points). The 
results of the global fit (with  $\sin 2\phi_1 = 0.99$)
are shown as solid and dashed curves, respectively.}
\label{belle_fig2}
\end{figure}

\begin{figure}[htb]
\centerline{\epsfysize 5.9 truein\epsfbox{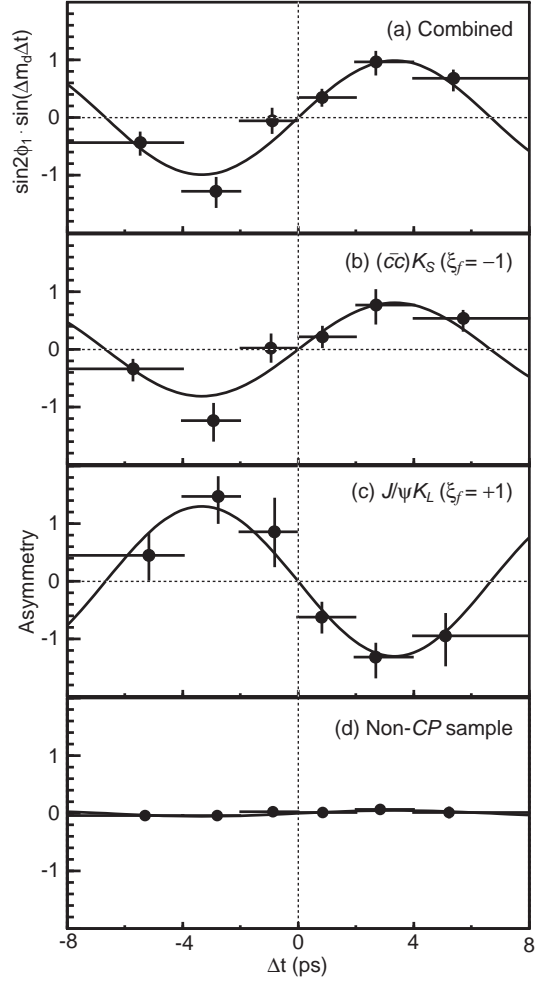}}
\caption{BELLE data on time dependent CP asymmetry.
The raw $\Delta$ distributions for $B^0$ and $\bar{B^0}$ tags.
(a) The asymmetry obtained
from separate fits to each $\Delta t$ bin for 
the full data sample; the curve is the result of 
the global fit. The
corresponding plots for the (b) $(c\bar{c})K_S$ ($\xi_f=-1$), (c) 
$J/\psi K_L$ ($\xi_f = +1$), and (d) $B^0$ control samples
are also shown.  The curves
are the results of the fit applied separately to the
individual data samples.
}
\label{belle_fig3}
\end{figure}

Examination of the time dependent distributions from the two
experiments (Figs.~\ref{babar_fig2},~\ref{belle_fig2},~\ref{belle_fig3}.)
shows clear indications of CP violation.
The raw time distributions for the $B^0$ and $\bar{B^0}$ samples are indeed
asymmetric. This is most striking in the BELLE figure.
Thus, CP violation is visible even in the raw time
distributions. 
The time dependent asymmetry distributions for the $\xi_f=\pm 1$
samples are indeed opposite as expected. Finally, the flavor tagged
samples show no time dependent asymmetry.

To extract the values of the CP violating parameter,
the two experiments perform unbinned maximum likelihood fits to the 
time distributions of the tagged and vertexed events.
The fits take into account the effects of background, vertex resolutions
and incorrect tagging.

In the summer of 2001, the first significant measurements
of the CP violating parameter $\sin 2 \phi_1$ were obtained
by BELLE and BABAR. BELLE finds
$$ \sin 2 \phi_1 = 0.99 \pm 0.14 \pm 0.06$$ with a statistical
significance of greater than six standard deviations\cite{belle_prl}. BABAR
finds $$ \sin 2 \phi_1 = 0.59 \pm 0.14 \pm 0.05$$ 
with a significance of 4.1 standard deviations\cite{babar_prl}.
The results
are based on data samples of comparable size (31 million and 32
million $B \bar{B}$ pairs, respectively). As discussed
above, the efficiencies and
resolutions of the two experiments are also quite comparable.
However, the two measurements are only marginally consistent. 
Larger data samples and additional more precise measurements
will be required to fully reconcile these two results.

\section{Determination of the other angles}

Experimental work on the determination of the other angles
$\phi_2$ and $\phi_3$ has just started. To obtain
precise results that test the validity of the Standard Model,
much larger data samples will be required. Below I briefly
describe some of the possible approaches.

To measure $\phi_2$ (a.k.a $\alpha$), the two most promising
approaches involve the use of the decay modes $B^0\to \pi^-\pi^+$
and $B^0\to \rho^{\pm}\pi^{\mp}$. The former is an example of 
CP eigenstate and is the most straightforward approach.
The interference in this mode between the direct decay and the decay via
mixing leads to a CP violating asymmetry as in the charmonium decay mode
$B^0\to \psi K_s$. 

However, there are several additional complications. 
The decay amplitude for $B^0\to \pi^+\pi^-$ contains a contribution
from a tree diagram ($b\to u \bar{u} d$) as well as a Cabibbo
suppressed penguin diagram ($b\to s \bar{u} s$). The penguin
contribution is not negligible and has a weak phase that is
different from the phase of the larger tree amplitude, which is
zero in the usual parameterization. 
Therefore the time dependent asymmetry, proportional to
$\sin(\Delta m \Delta t)$, 
which is measured is not equal to $\sin 2\phi_2$ but
instead will have a large unknown correction.
The presence of the extra contribution also induces an additional
time dependent term proportional to $\cos(\Delta m \Delta t)$. 
This is called penguin pollution.

There are a number of other purely experimental complications. The
branching fraction for the $B^0\to \pi^+\pi^-$ decay is somewhat
small. Using $22.6\times 10^6$ $B\bar{B}$ pairs BABAR finds 
${\cal B}(B^0\to \pi^+\pi^-)= (4.1\pm 1.0\pm 0.7)\times 10^{-6}$
\cite{babar_kpi}, 
while using a sample of $11.1\times 10^{6}$ $B\bar{B}$ pairs BELLE finds
${\cal B}(B^0\to \pi^+\pi^-)= (5.6\pm 2.3\pm 0.4)\times 10^{-6}$
\cite{belle_kpi}.  Moreover, even after 
the application of high momentum particle identification,
the CP eigenstate signal sits on a large continuum background.
The signal to continuum background ratio 
is a function of the tag method; this
effect must be taken into account in the CP extraction. The
conventional continuum suppression variables are also correlated 
with tagging. There is still residual
background from misidentified $B^0\to K^+\pi^-$ as well.

Despite these difficulties, the first attempts to
determine $\sin2\phi_2$ using the $\pi^+\pi^-$ mode 
have started. BABAR reports
results of a preliminary fit which gives 
$\sin 2\phi_2 (eff) = 0.03^{+0.53}_{-0.56}\pm 0.11$. BELLE expects
a sensitivity of $\pm 0.6$ and will present a result soon.
The size of the correction for penguin pollution in $\sin 2\phi_2
(eff)$ is unknown.

The angle $\phi_3$ is particularly difficult to measure.
There are at least three methods which give some information
on this angle.

The method which requires the least data but has the largest
theoretical uncertainty involves the measurement of the
branching fractions of the $K\pi$ and $\pi\pi$ decay modes.
The underlying idea is to compare decays where
 the interference of the penguin and tree amplitudes are different
to extract about information about the phase $\phi_3$.
For example, if $\phi_3>90^0$, then the interference term
is constructive for $B^0\to K^+\pi^-$ and destructive
for $B^0\to \pi^+\pi^-$, leading to a ratio for the widths
of the two modes greater than one. Conversely, if $\phi_3<90^0$
then a $K^-\pi^+/\pi^+\pi^-$ ratio less than unity is expected.
The current data somewhat favor $\phi_3>90^0$ whereas
the indirect CKM fits prefer $\phi_3\sim 60^0$. The interpretation
of this discrepancy is still unclear.

A theoretical industry has been developed to extract the
maximum possible information on $\phi_3$ from these and related
decay modes\cite{neubert_gamma}. 
However, it has been argued by some theorists
that other smaller diagrams
or final state interaction effects may invalidate 
these analyses\cite{sanda_gamma},\cite{ciuchini_gamma},
\cite{suzuki_gamma}.

Another method to determine $\phi_3$ uses the Cabibbo suppressed
decay mode $B^-\to D^0 K^-$\cite{dunietz_dk}. 
In the case that $D^0\to f_{CP}$, where
$f_{CP}$ is a CP eigenstate such as $K^- K^+, \pi^+\pi^-, K_s\pi^0$,
two amplitudes with different weak phases
will interfere. The asymmetry and rate again involve
the phase of $V_{ub}$. This method is theoretically cleaner 
than $B\to K\pi$ but
requires extremely large data samples. The current yield of 
$B^-\to D_{CP} K^-$ events from BELLE is $\sim 16$ events
in a 21 fb$^{-1}$ data sample\cite{belle_dk}. This yield may be increased
by about a
 factor of two by including additional CP eigenstates of the $D$ meson.
Another approach is to use doubly Cabibbo suppressed decays (DCSD)
of the $D$ meson. In this case, the event yield is much smaller
but the direct CP asymmetries are much larger\cite{dunietz_dk}.

Measurement of a time dependent
asymmetry in the mode $\bar{B}^0\to D^{*+}\pi^-$ also gives
information on the angle $\phi_3$. In this case, however, the asymmetry
is proportional to $\sin(2\phi_1 +\phi_3)$. The time dependence
is also more complicated than in the case of a simple CP eigenstate such
as $B^0\to \psi K_s$\cite{dunietz_dpi}. 
To reconstruct $\bar{B}^0\to D^{*+}\pi^-$,
one uses a partial reconstruction technique where the $\pi^-$ and
the low momentum pion from the $D^{*+}\to (D^0)\pi^+$ decay are observed but
the $D^0$ meson is not reconstructed. This increases the number
of useable signal candidates by roughly an order of magnitude.
Initial MC estimates indicate that, however, that even with the partial
reconstruction method, 200 fb$^{-1}$ 
(which corresponds to 17000 lepton tagged signal events) will give a precision
of $\pm 0.34$ for $\sin(2\phi_1+\phi_3)$ in this channel\cite{zheng}.

For the methods described above one usually measures
a trigonometric function of one of the angles such as
$\sin(2\phi_1)$. In such a case, there often exists
multiple solutions or ambiguities for the angle itself 
(e.g. four solutions
for $\phi_1$ from the initial $\sin(2\phi_1)$ measurements). 
One cannot simply discard the solutions
that are not consistent with the Standard Model. 
Instead to resolve these ambiguities, 
requires difficult high statistics measurements 
of quantities that depend on other functions of the angles such as
$\cos(\phi)$.

In addition to the program of measuring the angles of
the unitarity triangle, there is also the question of 
whether there are new CP violating phases from new interactions
or physics beyond
the Standard Model. At the moment, such new phases are
completely unconstrained. One way to attack this question is
to measure the time dependent CP asymmetry in penguin modes 
such as $B^0\to \phi K_s$ or $B^0\to \eta^{'} K_s$ 
and compare it to the asymmetry
in $B^0\to \psi K_s$. In the absence of new physics, 
they should be equal. However, if there are new physics 
contributions in penguin loops these asymmetries
will differ substantially.  Again an order of
magnitude more data is needed for 
stringent tests\cite{babar_phi},\cite{belle_etaks}.
This search for new physics
in CPV will be one of the most interesting aspects of
the next phase in B factory physics.


\vfill
\eject

\end{document}